\documentclass[aps,prd,showpacs%,twocolumn,
superscriptaddress]{revtex4}

\usepackage{graphicx}
\usepackage{dcolumn}
\usepackage{doi}
\usepackage{amsmath}
\usepackage{amssymb}
\usepackage{bm}
\usepackage{color}
\usepackage[normalem]{ulem}
\usepackage[dvipsnames]{xcolor}
\usepackage{hyperref}
\hypersetup{
colorlinks=true, % false: boxed links; true: colored links
linkcolor=blue,  % color of internal links
citecolor=cyan,  % color of links to bibliography
}

\begin{document}
\title{Axially symmetric and static solutions of Einstein equations with self-gravitating scalar field}
\author{Bobur Turimov}
\email{bturimov@astrin.uz}
\affiliation{Ulugh Beg Astronomical Institute, Astronomicheskaya 33,
Tashkent 100052, Uzbekistan }
\affiliation{Institute of Physics and Research Centre of Theoretical Physics and Astrophysics, Faculty of Philosophy and Science, Silesian University in Opava,
Bezru\v covo n\' am.13, CZ-74601 Opava, Czech Republic }
\author{Bobomurat Ahmedov}
\email{ahmedov@astrin.uz}
\affiliation{Ulugh Beg Astronomical Institute, Astronomicheskaya 33,
Tashkent 100052, Uzbekistan }
\affiliation{National University of Uzbekistan,
Tashkent 100174, Uzbekistan}
\author{Martin Kolo\v s}
\email{martin.kolos@fpf.slu.cz}
\affiliation{Institute of Physics and Research Centre of Theoretical Physics and Astrophysics, Faculty of Philosophy and Science, Silesian University in Opava,
Bezru\v covo n\' am.13, CZ-74601 Opava, Czech Republic }
\author{Zden\v ek Stuchl{\' i}k}
\email{zdenek.stuchlik@physics.cz}
\affiliation{Institute of Physics and Research Centre of Theoretical Physics and Astrophysics, Faculty of Philosophy and Science, Silesian University in Opava,
Bezru\v covo n\' am.13, CZ-74601 Opava, Czech Republic }
\date{\today}
\begin{abstract}
The exact axisymmetric and static solution of the
Einstein equations coupled to the axisymmetric and static gravitating scalar (or phantom) field is presented. The spacetimes modified by the 
scalar field are explicitly
given for the so-called $\gamma$-metric and the
Erez-Rosen metric with quadrupole moment $q$, and the
influence of the additional deformation
parameters $\gamma_*$ and $q_*$ generated by the scalar field is studied. It is shown that the null energy condition
is satisfied for the phantom field, but
it is not satisfied for the standard scalar field.
The test particle motion in both the
modified $\gamma$-metric and the Erez-Rosen quadrupole metric is studied; the circular geodesics are determined, and near-circular trajectories are explicitly presented for
characteristic values of the spacetime parameters.
It is also demonstrated that the parameters $\gamma_*$
and $q_*$ have no influence on the test particle
motion in the equatorial plane.
\end{abstract}

\maketitle

\section{Introduction}

One of the important problems in general relativity
is to find new exact analytical solutions of the Einstein field
equations. Numerous powerful methods have been
developed for the derivation of new solutions of the
gravitational field equations since Einstein discovered
the theory of general relativity in 1915. Well-known,
astrophysically relevant, external vacuum solutions
of Einstein field equations have been obtained by
Schwarzschild and Kerr for static and rotating
black holes, respectively.
In an early paper~\cite{Tolman39}, several static
solutions of the Einstein equations have been presented.
The Hartle-Thorne metric~\cite{Hartle67,Hartle68,Hartle69} describes the interior
and the vacuum spacetime outside any slowly
rotating astrophysical object as relativistic star. 
A huge number of interesting
exact solutions of the Einstein field equations can be
found in the books~\cite{Stephani03,Kramer80}.

Astronomical objects can be deformed for various reasons and consequently it is
interesting to study the spacetime of the
deformed compact gravitational objects.
In Ref.~\cite{Erez59} an exact axisymmetric
static vacuum solution of the Einstein equations in 
the case of a nonspherical mass distributed by a
compact object has been obtained. This solution is often called the quadrupolar (quadrupole moment) metric with an external $q$ mass quadrupole moment.
The exact solution of the Einstein equations for
deformed spacetime, which is called $\gamma$-metric,
has been obtained in Ref.~\cite{Janis68}
and later in~\cite{Voorhees70} it
has been derived in a different way.
These solutions  belong to the class of
Weyl type solutions and similar solutions have been studied by various authors, for instance in~\cite{Gutsunayev85,Quevedo85,
Quevedo86,Quevedo87,Quevedo89,Gutsunaev89,Quevedo90,Quevedo91b,
Frutos-Alfaro18,Boshkayev16,Contopoulos16}.

The geodesic motion of the test particles
in spacetime described by $\gamma$-metric
has been studied in~\cite{Chowdhury12} and
in the quadrupolar metric in~\cite{Quevedo89a}.

Recently, it has been shown that the massive scalar field may
give a much larger contribution to the gravitational field
around the slowly rotating neutron star in
comparison with that of the massless scalar field~\cite{Staykov18}.
The exact solution of the Einstein equations for the wormhole
with the scalar field has been recently studied~\cite{Gibbons17}.
Some approximate static solutions of the Einstein equations are shown
in~\cite{Fan15,Fan15a,Fan15b,Fan15c} under
the conformal field theory approach. The 
contribution of the scalar field
in the spacetime of static~\cite{Virbhadra97,Dadhich01,Herrera05,Herdeiro15,Erices15}
and rotating~\cite{Sen92} black holes have been
also studied. Regular and quintessential black hole solutions
has been recently extended to the rotating axially symmetric
ones using the Newman-Janis algorithm (NJA)~\cite{Toshmatov14,Toshmatov17}. 

In this paper, we are interested in getting
axisymmetric and static solutions of the
Einstein field equations taking into account the effect of  self
gravitating scalar field. For this reason, 
we first show derivation
of the Einstein field equations in Weyl and prolate
coordinates in close detail. Then after deriving the requested solution we test the effect of the
gravitating scalar field in the spacetime
of the static black hole using test particle motion.

The paper is organized in the following way.
In the Sec.~\ref{Solution} we provide in very detailed form
the exact analytical solutions of the Einstein
field equations with the self gravitating massless
scalar field obeying the Klein-Gordon equation.
For convenience in the calculations performance we use the
axisymmetric Weyl coordinates and
the prolate spheroidal coordinates.
We derive more general form of axisymmetric and static solutions of Einstein field equations 
including influence of the
external self-gravitating scalar field.
Section~\ref{gammametric} is the devoted to
derivation of the $\gamma$-metric solution, which
includes the external gravitating
scalar field. As a probe of the modified
$\gamma$-metric, we consider test particle motion
and the energy condition in its spacetime. In Sec.~\ref{quadrupolarmetric}
we obtain the modified quadrupolar metric
including the influence of the gravitating 
scalar field and then study particle motion 
in this spacetime. Finally,
in Sec.~\ref{Summary}
we summarize obtained results and
give future outlook related to the present work.

Throughout the paper we use a spacelike
signature $(-,+,+,+)$, a system of units
in which $G = c = 1$ and  restore them when we need to
compare the results with the observational data.
Greek indices are taken to run from $0$ to $3$,
Latin indices from $1$ to $3$.

\section{General solution of Einstein equations with self-gravitating scalar field\label{Solution}}

In this section we plan to incorporate into
the Einstein field equations the effect of a real massless self-gravitating
scalar field. The action for the system is described by the following form~\cite{Gibbons17}:
\begin{equation}\label{action}
{\cal S} = \int d^4x\sqrt{-g}\left(R - 2\epsilon g^{\mu\nu}
\partial_\mu\Phi \partial_\nu\Phi\right)\ ,
\end{equation}
where $g$ is the determinant of the
metric tensor $g_{\mu\nu}$ of the arbitrary
spacetime, $R$ is the Ricci scalar of the
curvature and $\Phi$ is the massless
gravitating scalar field, $\epsilon$
is the constant which is
responsible for the scalar field at $\epsilon = 1$
and the phantom field with value
$\epsilon = -1$, respectively.

Hereafter minimizing the action
in the equation (\ref{action})
one can obtain equations of motion of the
system which is described by the Einstein field
equations taking into account of
the gravitating scalar field and the
Klein-Gordon equation for the
gravitational scalar field in the following form
\begin{eqnarray}\label{eq1}
&& R_{\mu\nu} = 2\epsilon\, \partial_\mu\Phi
\,\partial_\nu\Phi\ ,
\\
\label{eq2}
&&
\square \Phi = 0 \ ,
\end{eqnarray}
where $R_{\mu\nu}$ is the Ricci tensor of the curvature
and $\square$ is the d'Alembertian in four dimensional
curved space. It is well known that the
equations (\ref{eq1})-(\ref{eq2})
are coupled differential equations and finding their
solutions is not easy task so far.
In this work we present axial-symmetric and
static solutions of the field equations
(\ref{eq1})-(\ref{eq2}) and compare the
solutions with those previously obtained in the literature.
%%%%%%%%%%%%%%%%%%%%%%%%%%%%%%%%%%%%%%%%%%%%%

\subsection{Axisymmetric and static solution}

In order to simplify the problem we
can make an assumption that the gravitating
scalar field is axially symmetric and
stationary. In Weyl coordinates
($t,\rho,\phi,z$) the general form
of the static metric can be described by
\begin{equation}\label{Weylmetric}
ds^2 = - e^{2U} dt^2 + e^{-2U}\left[e^{2V}
(d\rho^2+dz^2) +\rho^2 d\phi^2\right]\ ,
\end{equation}
where $U$ and $V$ are the functions of the
coordinates $\rho$ and $z$, respectively.
Then the explicit form of the field equations
(\ref{eq1})-(\ref{eq2}) for the spacetime
metric (\ref{Weylmetric}) can be
written as~\cite{Gibbons17}
\begin{eqnarray}\label{FieldEquation1}
&&
\Delta\Phi=\Phi_{\rho\rho}+\frac{1}{\rho}\Phi_{\rho}+\Phi_{zz}=0\ ,
\\
&&
\Delta U=U_{\rho\rho}+\frac{1}{\rho}U_{\rho}+U_{zz}=0\ ,
\\
&&
V_\rho = \rho\left(U_\rho^2-U_z^2+
\epsilon\Phi_\rho^2-\epsilon\Phi_z^2\right)\ ,
\\
&&
V_z = 2\rho\left(U_\rho U_z+\epsilon \Phi_\rho\Phi_z\right) \ ,
\label{FieldEquation4}
\end{eqnarray}
where subindices indicate the derivative with
respect to the coordinates $\rho$ and $z$, respectively.

For the convenience one can consider
the prolate coordinates
($t,X,Y,\phi$) in which the spacetime metric (\ref{Weylmetric})
can be rewritten in the following form~\cite{Quevedo85,Quevedo86,Quevedo87,Quevedo89}:
\begin{eqnarray}\label{metric}
ds^2 = - e^{2U} dt^2 + \sigma^2 e^{-2U}
\left[e^{2V}\left(X^2-Y^2\right) 
%\right.\\\nonumber\left.\times
\left(\frac{dX^2}{X^2-1}+\frac{dY^2}{1-Y^2}\right)
+(X^2-1)(1-Y^2) d\phi^2\right]\ ,
\end{eqnarray}
where $\sigma$ is the dimensional parameter,
later in the text the physical meaning of
this parameter will be introduced.

Here we can introduce useful notations which are
the relations between the prolate spheroidal
coordinates ($X,Y,\phi$) and Weyl coordinates
($\rho, z,\phi$) indicated as
\begin{equation}
\rho = \sigma\sqrt{(X^2-1)(1-Y^2)}\ , \,\,\, z = \sigma XY \ ,
\,\,\, \phi = \phi\ ,
\end{equation}
and similarly they can be related
with the spherical coordinates ($r,\theta,\phi$) in the following
form
\begin{equation}\label{Sphere}
X = \frac{r}{\sigma}-1 \ ,\quad Y = \cos\theta \ ,
\quad \phi = \phi\ .
\end{equation}
Note that here zeroth (temporal)
component of the coordinate $t$ is the
same in all these coordinates.

Finally, the field equations
(\ref{FieldEquation1})-(\ref{FieldEquation4})
can be rewritten in terms of prolate coordinates
$X$ and $Y$ in the form
\begin{eqnarray}
\label{Equation1}
&&
\left[\left(X^2-1\right)\Phi_X\right]_X +
\left[\left(1-Y^2\right)\Phi_Y\right]_Y=0\ ,
\\
\label{Equation2}
&&
\left[\left(X^2-1\right)U_X\right]_X +
\left[\left(1-Y^2\right)U_Y\right]_Y=0\ ,
\\
\label{Equation3}
%\nonumber
&& V_X = \frac{1-Y^2}{X^2-Y^2}\Big[X(X^2-1)U_X^2
%\\\nonumber &-&
-
X(1-Y^2)U_Y^2-2Y(X^2-1)U_XU_Y\Big]
%\\&+&
+
(U\to\epsilon\Phi) \ ,
\\
\label{Equation4}
%\nonumber
&&V_Y = \frac{X^2-1}{X^2-Y^2}\Big[Y(X^2-1)U_X^2
%\\&-&
-Y(1-Y^2)U_Y^2
+2X(1-Y^2)U_XU_Y\Big]
%\\&+&
+
(U\to\epsilon\Phi) \ .
\end{eqnarray}

One can easily see that the equations (\ref{Equation1})
and (\ref{Equation2}) are similar to each other,
one can seek their solutions in the following separable
form $\{\Phi,U\} = f(X)g(Y)$ (See e.g.,~\cite{Quevedo85})
and using the equations (\ref{Equation1}) and
(\ref{Equation2}) one can write the following
Legendre equations for the
functions $f(X)$ and $g(Y)$ in the form
\begin{eqnarray}\label{Eqf}
&&
\left[\left(X^2-1\right)f_X\right]_X - l(l+1)f=0 \ ,
\\
&&
\left[\left(1-Y^2\right)g_Y\right]_Y + l(l+1)g=0 \ ,
\label{Eqg}
\end{eqnarray}
where $l$ is the multipole number that
can take the integer values. The solutions
of the equations (\ref{Eqf}) and (\ref{Eqg}) are
\begin{eqnarray}
\label{Solf}
f(X) &=& C_{1l} P_l(X) + C_{2l} Q_l(X) \ ,
\\
g(Y) &=& C_{3l} P_l(Y) + C_{4l} Q_l(Y) \ ,
\label{Solg}
\end{eqnarray}
where $P_l(X)$ is the Legendre polynomial,
$Q_l(Y)$ is the Legendre function of the
second kind and $C_{1l}-C_{4l}$ are the
integration constants, respectively.
From the physical point of view both
solutions $\{\Phi,U\}$ should be
asymptotically flat which means
\begin{equation}
\lim_{X\to \infty} f(X) = 0 \ , \qquad C_{1l}=0 \ ,
\end{equation}
and they should be regular everywhere
\begin{equation}
\lim_{Y\to 0}g(Y) = {\rm const} \ ,\quad C_{4l}=0 \ .
\end{equation}
In order to get correct results,
here we use the following property of the Legendre
function of the second kind
$Q_l(-X) = (-1)^{l+1}Q_l(X)$.
Consequently, the solutions for the functions
$\Phi(X,Y)$ and $U(X,Y)$ can be written as
\begin{eqnarray}
\label{SolF}
&&
\Phi(X,Y) = \sum_{l=0}^{\infty}(-1)^{l+1}p_l Q_l(X)P_l(Y) \ ,
\\
\label{SolU}
&&
U(X,Y) = \sum_{l=0}^{\infty}(-1)^{l+1}q_l Q_l(Y)P_l(Y) \ ,
\end{eqnarray}
{where new constants (of integration) $q_l$ and $p_l$ are  
	the standard multipole moments of the gravitational compact object and 
	the multipole moments generated by the gravitating scalar field, respectively, 
	and they are totally independent quantities.}
The unknown function $V(X,Y)$ can be
found by solving the equations
(\ref{Equation3}) and (\ref{Equation4}) which is quite
a complicated task. The explicit form of the function
$V(X,Y)$ is given by (detailed calculation are shown
in Ref.~\cite{Quevedo89})
\begin{equation}\label{SolV}
V(X,Y) = \sum_{l,n=0}^{\infty}
(-1)^{l+n} (q_lq_n+\epsilon p_lp_n)
\Gamma^{\{ln\}} \ ,
\end{equation}
where exact form of $\Gamma^{\{ln\}}$ can be found
in the Appendix~\ref{App1}.

In order to find the physically meaningful
solution one can set $\epsilon=0$, $q_0=1$ and
$q_l=0$ ($l>0$) in solutions (\ref{SolU})
and (\ref{SolV}) and obtain the well-known
Schwarzschild solution
\begin{equation}
U = \frac{1}{2}\ln\frac{X-1}{X+1} = \frac{1}{2}
\ln\left(1-\frac{2\sigma}{r}\right) \ ,
\end{equation}
\begin{equation}
V = \frac{1}{2}\ln\frac{X^2-1}{X^2-Y^2}
=\frac{1}{2}\ln\frac{r^2-2\sigma r}{r^2-2\sigma r+\sigma^2\sin^2\theta} \ .
\end{equation}
Here one can easily see that the dimensional
parameter $\sigma$ is the total mass
of the compact object $\sigma=M$.

\section{Analytic solution of the Einstein equations with self-gravitating scalar field for the $\gamma$-metric\label{gammametric}}

In this section we study the zeroth order
approximation of the solutions for the
profile functions and  present
the generalized form of  the well-known 
$\gamma$-metric  including the effect
of the scalar field. By considering the case when
$q_0=\gamma$, $p_0=\gamma_*$ and
$q_l=p_l=0$ $(l>0)$ in the equations
(\ref{SolF})-(\ref{SolV}) one can obtain the
results for the functions $\Phi(X,Y)$, $U(X,Y)$
and $V(X,Y)$ in the following form
\begin{eqnarray}
\label{Phiquad}
\Phi(X,Y) &=& \frac{\gamma_*}{2}\ln\frac{X-1}{X+1} \ ,
\\
\label{Uquad}
U(X,Y) &=& \frac{\gamma}{2}\ln\frac{X-1}{X+1} \ ,
\\
\label{Vquad}
V(X,Y) &=& \frac{\gamma^2+\epsilon \gamma_*^2}{2}
\ln\frac{X^2-1}{X^2-Y^2} \ ,
\end{eqnarray}
Using the coordinate transformation
in the expression (\ref{Sphere}) we can obtain the
generalized form of the $\gamma$-metric
in spherical coordinates
\begin{widetext}
\begin{eqnarray}\label{metric1}
\nonumber
ds^2 =& -& \left(1-\frac{2M}{r}\right)^{\gamma}
dt^2
\\
&+& \left(1-\frac{2M}{r}
\right)^{1-\gamma}
\left\{\left(1-\frac{M^2\sin^2\theta}{r^2-2Mr}
\right)^{1-\gamma^2-\epsilon \gamma_*^{2}}
\left[\left(1-\frac{2M}{r}\right)^{-1}dr^2
+r^2d\theta^2\right]
+r^2\sin^2\theta d\phi^2\right\}\ ,
\end{eqnarray}
\end{widetext}
and the scalar field has a form
\begin{equation}\label{Fi}
\Phi(r)=\frac{\gamma_*^2}{2}\ln\left(1-\frac{2M}{r}\right) \ .
\end{equation}
Note that in the absence of the scalar field, 
which is when $\epsilon=0$,
we get the $\gamma$-metric. From the equation (\ref{metric1})
one can see that the scalar field contributes into $g_{rr}$
and $g_{\theta\theta}$ components of the metric tensor.
Other two components of the metric tensor
do not depend on the parameter $\gamma_*$ produced by the gravitating scalar field.

\begin{figure*}[h]%[htbp]
\begin{center}
\includegraphics[width=1\textwidth]{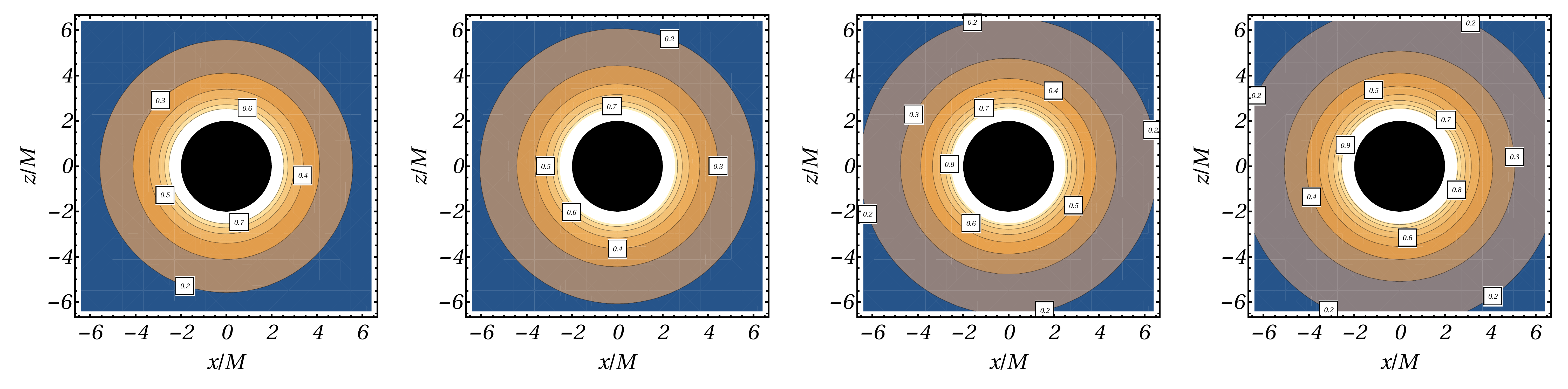}
\end{center}
\caption{ The shape of the scalar field
$\Phi(r,\theta)$ described by the equation (\ref{Fi})
in $x-z$ plane for the different values of
$\gamma_*$ parameter:
$\gamma_*=0.9$, $\gamma_*=1$, $\gamma_*=1.1$ and
$\gamma_*=1.2$.\label{Sfield}}
\end{figure*}

In the expression (\ref{Fi}) we can see that
the scalar function $\Phi(r)$ depends on the
radial coordinate only. Figure~\ref{Sfield}
draws the equipotential surface of the gravitating
scalar field $\Phi(r)$ in the ($x-z$) plane for
the different values of the $\gamma_*$ parameter.
One can easily see that with increasing the $\gamma_*$
parameter the gravitational force is getting
stronger and the spacetime around the object
will be deformed due to the presence of the
scalar field as shown in Fig.~\ref{Sfield}.

\subsection{Energy conditions\label{EnerCon}}
In this subsection we briefly study energy condition
in the spacetime of the generalized $\gamma$-metric
 given in the equation (\ref{metric1}).
The energy-momentum tensor for the
scalar field can be expressed as
\begin{equation}\label{EMtensor}
T_{\mu\nu} = \epsilon\left(\partial_\mu\Phi\partial_\nu\Phi
-\frac{1}{2}g_{\mu\nu}g^{\alpha\beta}\partial_\alpha
\Phi\partial_\beta\Phi\right)\ ,
\end{equation}
from the expression (\ref{EMtensor})
the energy density and the components of
the pressure can be defined as
$\rho = T_0^0$ and $P_i = T_i^i$, and
the explicit form of the energy density
and the components of the pressure is
\begin{eqnarray}\label{EMtensor1}
\nonumber
\rho = P_\theta = P_\phi = -P_r
%\\\nonumber
&=&
-\frac{\epsilon\gamma_*^2 M^2}{2r^4}
\left(1-\frac{2M}{r}\right)^{\gamma -2}
%\\&\times &
\left(1-\frac{M^2\sin^2\theta}
{r^2-2Mr}
\right)^{\gamma^2+\epsilon\gamma_*^2-1} \ .
\end{eqnarray}

\begin{figure*}[htbp]
\begin{center}
\includegraphics[width=1\textwidth]{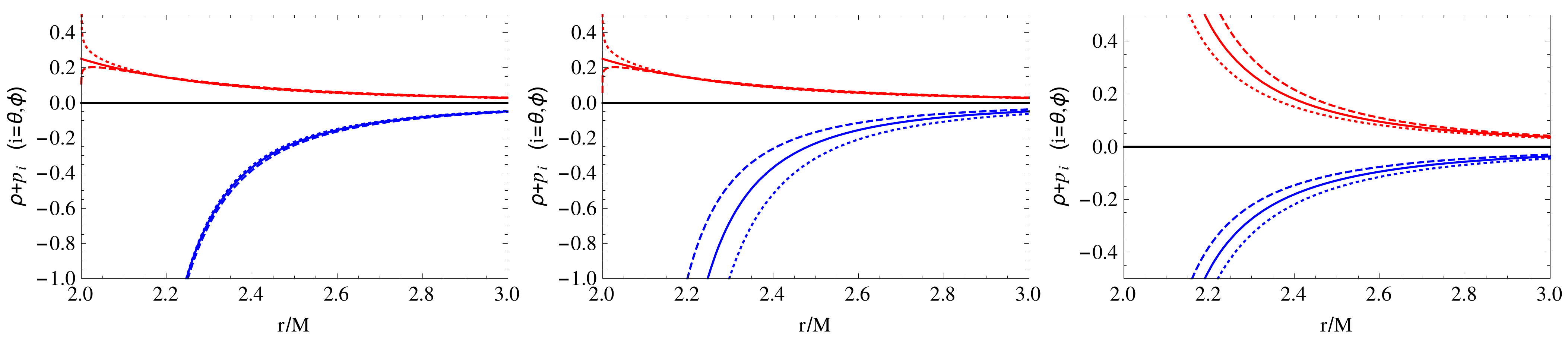}
\end{center}
\caption{{Radial dependence of
	$\{\rho + P_i\}$ ($i=r,\theta,\phi$)
	for the different values of the parameters
	$\gamma$ and $\gamma_*$. 
	(Left panel) Solid line corresponds to $\gamma=1$, dashed line to $\gamma=0.9$ and dashed line to $\gamma=1.1$ at $\gamma_* =1$ and $\theta=\pi/2$.
	(Central panel) Solid line corresponds to $\gamma_*=1$, dashed line to $\gamma_*=0.9$ and dashed line to $\gamma_*=1.1$ at $\gamma =1$ and $\theta=\pi/2$.
	(Right panel) Solid line corresponds to $\gamma=1$, dashed line to $\gamma=0.9$ and dashed line to $\gamma=1.1$ at $\gamma_* =1$ and $\theta=0$.}
	\label{EMTensor}}
\end{figure*}

The null energy condition (NEC)
can be found from the expression $\rho+P_i\ge 0$
($i=r,\theta,\phi$),
using the equation (\ref{EMtensor1}) as
\begin{eqnarray}
\label{rhor}
\rho + P_r & \equiv & 0\ ,
\\%\nonumber
\label{rhotf}
\rho + P_\theta &=& \rho + P_\phi =
-\frac{\epsilon\gamma_*^2 M^2}{r^4}
\left(1-\frac{2M}{r}\right)^{\gamma -2}
%\\&\times &
\left(1-\frac{M^2\sin^2\theta}
{r^2-2Mr}
\right)^{\gamma^2+\epsilon\gamma_*^2-1}\ .
\end{eqnarray}
{The physical interpretation of NEC 
	is that the energy density measured by an observer 
	traversing along null curve is always positive (never negative). 
	One can see that the expression (\ref{rhor}) 
	is always satisfied by the NEC condition for 
	the spacetime metric (\ref{metric1}) while the 
	expression (\ref{rhotf}) satisfies the NEC condition
	only in the case when $\epsilon \le 0$ 	which corresponds to the phantom field. This means that the 
    observer traversing along null curve can measure positive energy even in the case of the antigravitating phantom  
    scalar field. Figure~\ref{EMTensor} shows the NEC precisely where the radial 
	dependence of $\rho+P_i$ ($i=r,\theta,\phi$). }
    %-------------------------------------------------------------------------
%-------------------------------------------------------------------------%

\subsection{Test particle motion\label{ParMot}}
In this subsection we study test particle motion
in the spacetime metric (\ref{metric1}).
The Hamiltonian for test particle
with mass $m$ can be
written in the form~\cite{Kolos15}
\begin{equation}\label{Hamiltonian}
H =  \frac{1}{2} g^{\mu\nu}
p_\mu p_\nu + \frac{1}{2} \, m^2 \ ,
\end{equation}
where $p^\mu = m u^\mu$ is the kinematical
four-momentum.
The equations for particle motion are
\begin{equation}
\frac{d x^\mu}{d\zeta} \equiv p^\mu =
\frac{\partial H}{\partial p_\mu} \ ,
\qquad
\frac{d p_\mu}{d\zeta} = - \frac{\partial H}{\partial x^\mu}\ .
\end{equation}
Here the affine parameter $\zeta$ of
the particle is related to
its proper time $\tau$ by the
relation $\zeta=\tau/m$.

Because of the symmetries of the modified
$\gamma$-metric spacetime (\ref{metric1})
one can easily find the
conserved quantities that are the energy
and the axial angular momentum of
the particle and can be expressed as
\begin{eqnarray}
E &=& - p_t = m g_{tt} \frac{dt}{d\tau}\ ,
\\
L &=& p_\phi = m g_{\phi\phi} \frac{d\phi}{d\tau} \ .
\end{eqnarray}
Introducing for convenience the specific parameters, energy ${\cal E}$ and axial angular momentum ${\cal L}$
\begin{equation}
{\cal E} = \frac{E}{m}\ , \quad {\cal L} = \frac{L}{m}\ ,
\end{equation}
one can rewrite the Hamiltonian
(\ref{Hamiltonian}) in the form
\begin{equation}
H = \frac{1}{2} g^{rr} p_r^2 + \frac{1}{2} 
g^{\theta\theta} p_\theta^2  + \frac{m^2}{2} g^{tt} \left[{\cal E}^2 - V_{\rm eff}(r,\theta)\right]\ , \label{HamHam}
\end{equation}
where $V_{\rm eff}(r,\theta)$ denotes the effective potential of the test particle
which is given by the relation
\begin{eqnarray}\label{Veff}
V_{\rm eff}(r,\theta) &\equiv &
-
g_{tt}\left(1+g^{\phi\phi}{\cal L}^2\right)
\\\nonumber
&=& \left( 1-\frac{2M}{r} \right)^\gamma \left[ 1+\frac{{\cal L}^2}{r^2 \sin^2\theta}\left( 1-\frac{2M}{r} \right)^{\gamma-1}\right]\ .
\end{eqnarray}
It is important to note, that the effective potential $V_{\rm eff}(r,\theta)$ depends only on the metric parameter $\gamma$ while it is free of the parameters $\epsilon$ and $\gamma_*$.

The particle motion is limited by the energetic
boundaries given by
\begin{eqnarray}
{\cal E}^2 = V_{\rm eff} (r,\theta)\ . \label{MotLim}
\end{eqnarray}

\begin{figure*}
\includegraphics[width=\hsize]{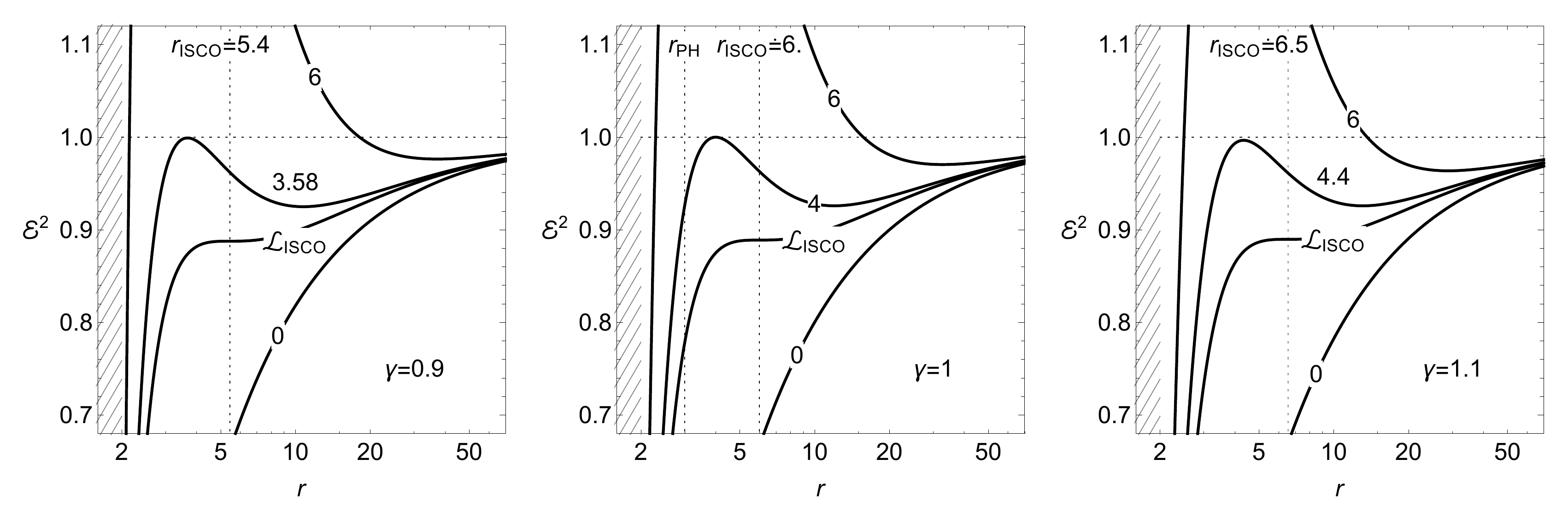}
\caption{Radial profiles of effective potential
in equatorial plane $V_{\rm eff}(r,\pi/2)$ for
the various values of angular momentum $L$.
In the plots the different values for $\gamma$ parameter are used.\label{figVeff}}
\end{figure*}

\begin{figure*}
\includegraphics[width=\hsize]{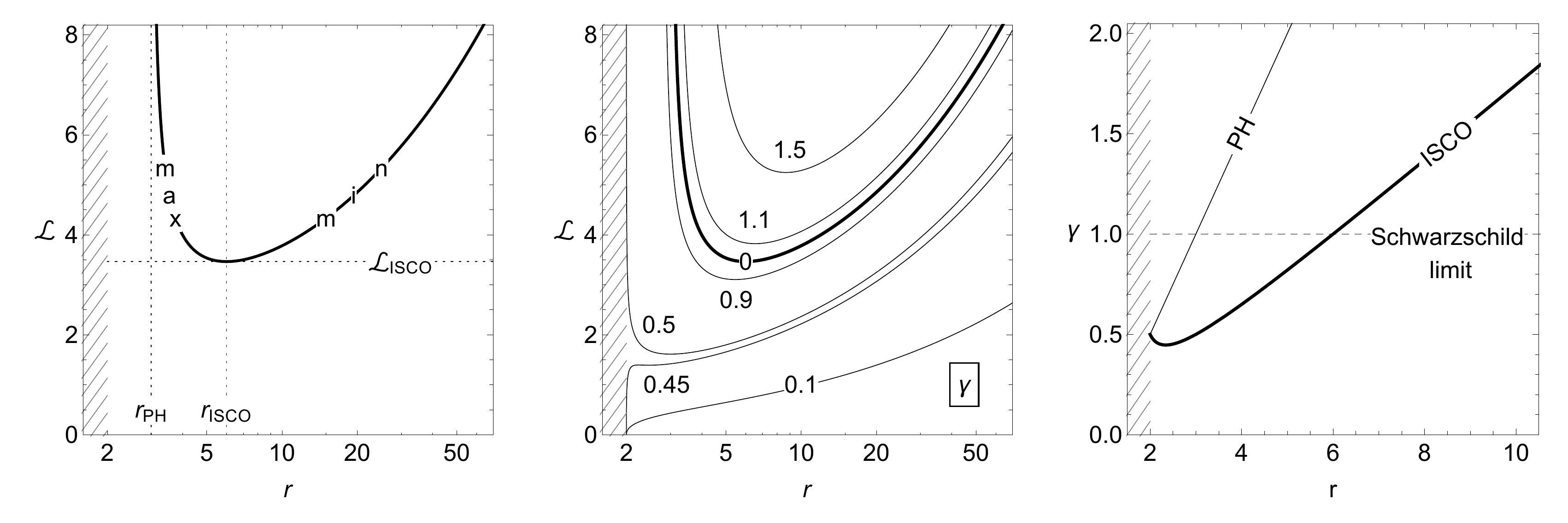}
\caption{ (Left panel) Position of extrema 
(max. min.) of the effective potential, giving stable (min) and unstable (max) circular orbits for the Schwarzschild ($\gamma=1$) spacetime.
(Central panel) Position of extrema (max. min.) of effective potential for the different values 
of the $\gamma$ parameter.
(Right panel) Position of the ISCO and photon orbit in the dependence from the parameter the $\gamma$. \label{figLcirc}}
\end{figure*}

Now we properly investigate the features of the
effective potential (\ref{Veff}) represented
in Fig.~\ref{figVeff}.  The stationary points
of the effective potential $V_{\rm eff}(r,\theta)$
function, where maxima or minima can exist,
are given by the equations
\begin{eqnarray}
\partial_r V_{\rm eff}(r,\theta) = 0\ ,
\qquad
\partial_\theta V_{\rm eff}(r,\theta) = 0 \ .  \label{extrem}
\end{eqnarray}
The second of the extrema equations
(\ref{extrem}) gives $\theta=\pi/2$. In other
words, all extrema of the $V_{\rm eff}(r,\theta)$
function are located at the equatorial plane and
there is no off-equatorial extrema. The first
extrema equation of (\ref{extrem}) leads to
equation being quadratic with respect to the specific
angular momentum ${\cal L}$
and hence the circular orbits can be
determined by the relation~\cite{Chowdhury12}
\begin{equation}
{\cal L}^2 = {\cal L}_{\rm ext}^2(r)
 \equiv \frac{\gamma Mr^2}{r-M(1+2\gamma)}
 \left(1-\frac{2 M}{r}\right)^{1-\gamma}\ . \label{Lcirc}
\end{equation}
At Fig.~\ref{figLcirc} the function
${\cal L}_{\rm ext}(r)$ is plotted for various values of parameter $\gamma$. Similarly,
the energy of {the test particle} 
can be expressed as
\begin{equation}
{\cal E}^2 = {\cal E}_{\rm ext}^2(r)
 \equiv \frac{r-M(1+\gamma)}{r-M(1+2\gamma)}
 \left(1-\frac{2 M}{r}\right)^{\gamma}\ . \label{Ecirc}
\end{equation}

The local extrema of  ${\cal L}_{\rm ext}(r)$ function is equivalent to $\partial^{2}_{r} V_{\rm eff}(r,\theta=\pi/2) = 0$ condition and they determine the innermost stable circular orbits (ISCO) radius located at
\begin{equation}\label{ISCO}
r_{\rm ISCO}/M=1+3\gamma +\sqrt{5\gamma^2-1}\ ,
\end{equation}
and from equation (\ref{ISCO}) we can find that $\gamma$
parameter should be $\gamma \ge 1/\sqrt{5}$.

The unstable circular photon orbit $m=0$
given by the divergence of the effective potential
(\ref{Veff}) will be located at
\begin{equation}
r_{\rm ph}/M = 1 + 2 \gamma\ .
\end{equation}
In the case when $\gamma=1$ one can have
$r_{\rm ISCO} = 6M$ and $r_{\rm ph}=3M$ which are
responsible for the radius of the ISCO and photon sphere, respectively,
in the Schwarzschild spacetime.

In Fig.~\ref{figLcirc} the various dependences 
of the radius of the ISCO and the photon sphere 
are shown. In the range of the values of 
the $\gamma\ge 1$ one can see that
with increasing the $\gamma$ parameter the radius of
ISCO and photon sphere increase while in the range of the values $1/\sqrt{5}\le\gamma\le 1$ they are small in comparison with that in general relativity.

\begin{figure*}
\includegraphics[width=\hsize]{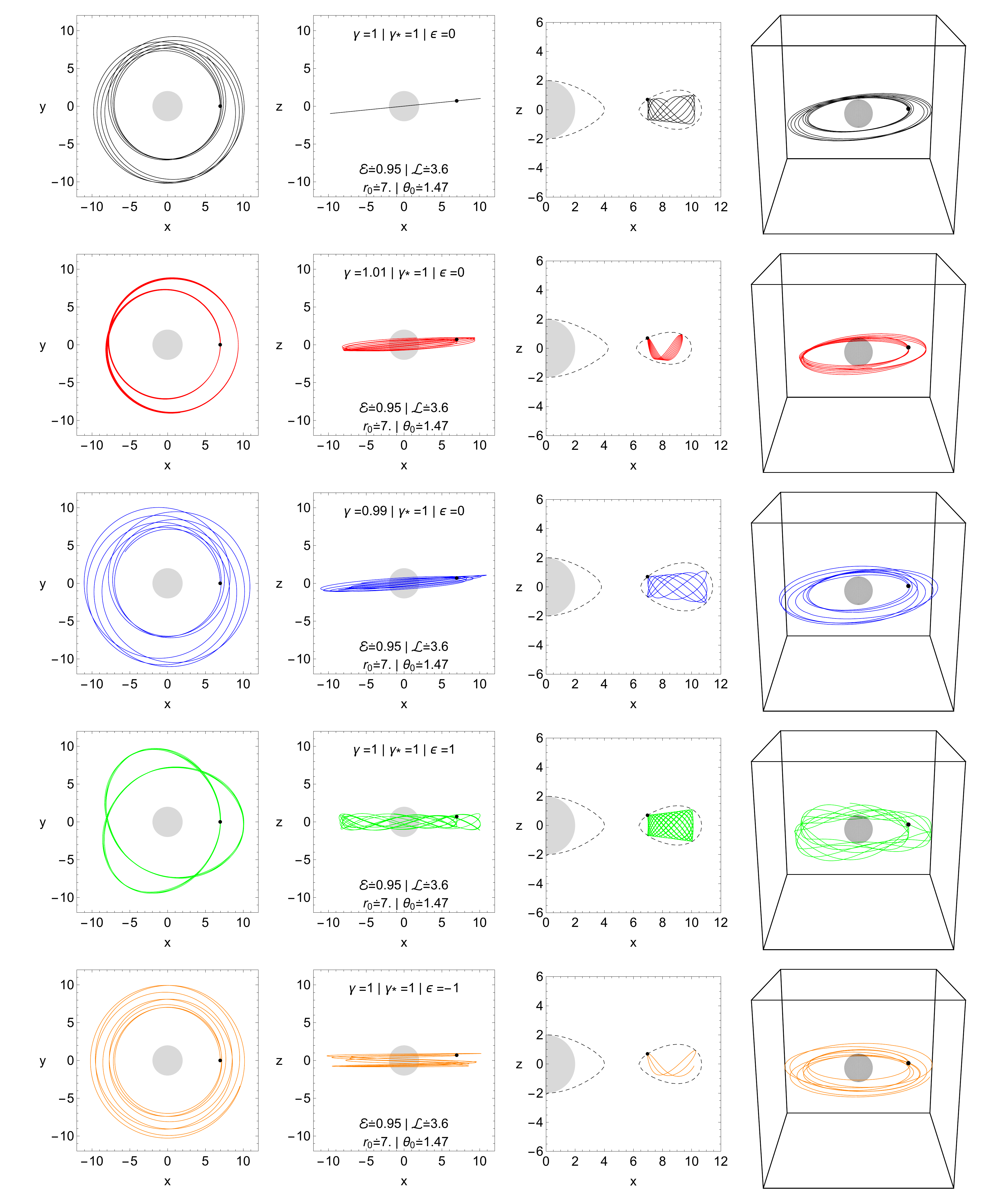}
\caption{{Test particle trajectories in the spacetime metric (\ref{metric1}) for the different values of parameters $\gamma$, $\gamma_*$ and $\epsilon$. 
In first and second (including third) columns
particle trajectories $x-y$ and $x-z$ planes are given
while in the fourth column a 3D $x-y-z$ pattern of 
particle trajectory is shown.}\label{Trajectory}}
\end{figure*}

{One can easily see that the 
Eqs. (\ref{Lcirc}), (\ref{Ecirc}) and (\ref{ISCO}) for  the angular momentum, the energy     and radius of ISCO of the test particle, respectively,  
do not contain $q$ which means that the gravitating 
    scalar field does not act on the test particles in the equatorial 
    plane. Numerical calculations show that the effects of the 
    gravitating scalar field can be seen in particle motion
    in off-equatorial plane. As a test of the spacetime geometry 
    (\ref{metric1}) we have presented  the 
    particle trajectories for the different values 
    of the metric parameters $\gamma$, $\gamma_*$ and $\epsilon$ 
    in several planes in Fig.~\ref{Trajectory}.}

%-------------------------------------------------------------------------%
%-------------------------------------------------------------------------%

%-------------------------------------------------------------------------%
%-------------------------------------------------------------------------%
%------------------------------------------------------------------------
%-------------------------------------------------------------------------

\section{Analytic solution of the Einstein equations with self-gravitating scalar field for the quadrupole-moment metric\label{quadrupolarmetric}}
In this section we briefly consider the
influence of the gravitating scalar field
into the quadrupole moment metric, which is described
by Erez-Rosen~\citep{Erez59}, with two free
parameters of the black hole, mass $M$ and
mass quadrupole moment $q$.
Now we can consider the next leading order approximation
in the coefficients $q_l$ and $p_l$
in the solutions (\ref{SolF})-(\ref{SolV})
of the field equations.
We study the case when $q_0=p_0=1$,
$q_1=p_1=0$
(which is not existed of the mass dipole moment),
and $q_l=p_l=0$ ($l>2$), and then we  have
\begin{widetext}
\begin{eqnarray}
\label{Phiq}
\Phi(X,Y) &=& \frac{1}{2}\ln\frac{X-1}{X+1}
%\\\nonumber &+&
+\frac{q_*}{2}(3Y^2-1)
\left(\frac{3X}{2}+\frac{3X^2-1}{4}\ln\frac{X-1}{X+1}\right) \ ,
\\
\label{Uq}
U(X,Y) &=& \frac{1}{2}\ln\frac{X-1}{X+1}
%\\\nonumber &+&
+\frac{q}{2}(3Y^2-1)
\left(\frac{3X}{2}+\frac{3X^2-1}{4}\ln\frac{X-1}{X+1}\right) \ ,
\\
\nonumber
\label{Vq}
V(X,Y) &=& \frac{(1+q)^2+\epsilon(1+q_*)^2}{2}
\ln\frac{X^2-1}{X^2-Y^2}
%\\\nonumber &-&
-\frac{3(q+\epsilon q_*)}{2}(1-Y^2)
\left(X\ln\frac{X-1}{X+1}+2\right)
\\\nonumber
&+&\frac{9(q^2+\epsilon q_*^{2})}{16}(1-Y^2)
\Big[X^2+Y^2-9X^2Y^2-\frac{4}{3}
%\\\nonumber &+&
+X\left(X^2+7Y^2-9X^2Y^2-\frac{5}{3}\right)
\ln\frac{X^2-1}{X^2-Y^2}
\\
&+&
\frac{1}{4}\left(X^2-1\right)
\left(X^2+Y^2-9X^2Y^2-1\right)
\ln^2\frac{X^2-1}{X^2-Y^2}
\Big] \ ,
\end{eqnarray}
\end{widetext}
where $q$ and $q_*$ are the mass quadrupole
moments of the gravitational object.
The Erez-Rosen solution~\cite{Erez59}
can be obtained in the limiting case when $q_*=0$.
In order to find the physically meaningful
solution for the scalar field
one writes it in terms of the spherical
coordinates in the form
\begin{eqnarray}\label{Phisphere}
\Phi(r,\theta) &=& \frac{1}{2}
\ln\left(1-\frac{2M}{r}\right)
+
\frac{q_*}{2}
\left[\frac{3r^2-6Mr+2M^2}{4M^2}
\ln\left(1-\frac{2M}{r}\right)
+
\frac{3(r-M)}{2M}\right] (3\cos^2\theta-1) \ ,
\end{eqnarray}
and in the weak field approximation
the equation (\ref{Phisphere}) has a form
\begin{equation}\label{WeakFieldPhi}
\Phi(r,\theta) \simeq - \frac{M}{r}
+\frac{q_*M^3}{15 r^3}(3\cos^2\theta-1) \ .
\end{equation}
We can see that the first linear term in the right hand side
of the equation (\ref{WeakFieldPhi})
is responsible for Newtonian potential,
the second term is responsible
for the quadrupole moment potential, where
$q_*$ is dimensionless mass quadrupole moment
produced by the gravitating scalar field.

\begin{figure*}[htbp]
\begin{center}
\includegraphics[width=1.0\textwidth]{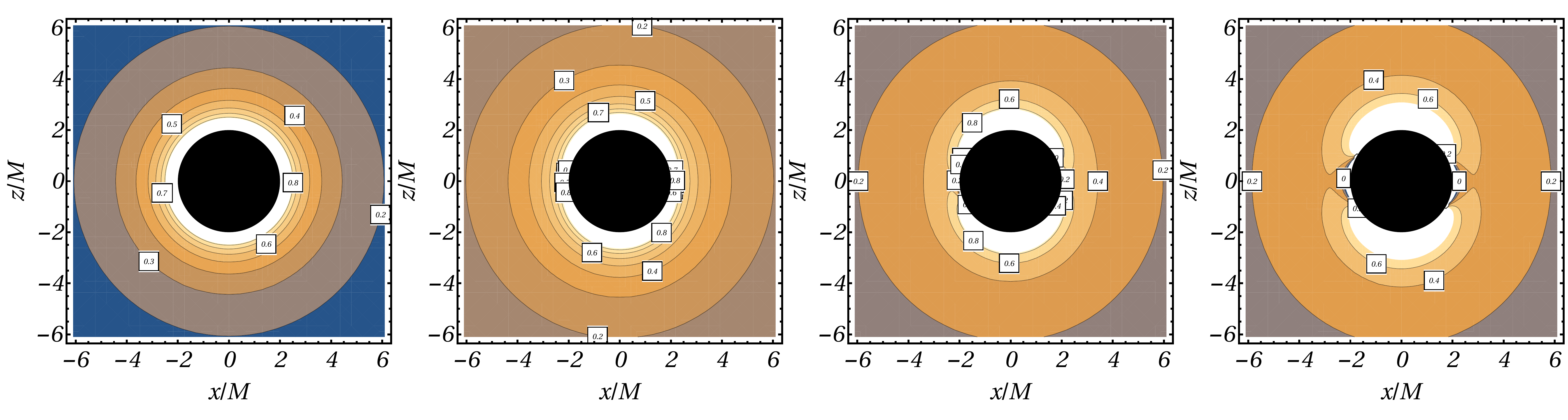}
\end{center}
\caption{ The equipotential surface of the
scalar potential $\Phi(r,\theta)$ in
$x-z$ plane
for the different values of the mass quadrupole
moment: $q_*=0$,
$q_*=0.2$, $q_*=0.5$ and $q_*=1$. \label{ScalarPotential}}
\end{figure*}

In Fig.\ref{ScalarPotential} the equipotential 
surface of the scalar field
$\Phi(r,\theta)$ using the expression (\ref{Phisphere})
for the different values of the
quadrupole moment $q_*$ is illustrated. One can
easily see that due to the $q_*$
parameter the spacetime around the
black hole is axially deformed.

It is a convenient to write simple analytical 
form of the metric for further calculations.
In the linear approximation of
the mass quadrupole moments
$q$ and $q_*$, one can write the
following spacetime metric
\begin{widetext}
\begin{subequations}\label{metric2}
\begin{eqnarray}
\label{g00}
g_{tt} &=& -\left(1-\frac{2M}{r}\right)
\Big[1+qF_1(r)P_2(\cos\theta)\Big]
+
{\cal O}(q^2)\ ,
\\
\label{grr}
\nonumber
g_{rr} &=& \left(1-\frac{2M}{r}\right)^{-1}
\left(1+\frac{M^2\sin^2\theta}{r^2-2 M r}\right)^{-\epsilon}
\\
&\times &
\left\{1+qF_1(r)
P_2(\cos\theta)-(q+\epsilon q_*)\left[
2\ln\left(1+\frac{M^2\sin^2\theta}{r^2-2 M r}\right)
+
3F_2(r)\sin^2\theta\right]\right\}
+
{\cal O}(q^2,q_*^2)\ ,
\\
\label{gtt}
\nonumber
g_{\theta\theta} &=&
r^2\left(1+\frac{M^2\sin^2\theta}{r^2-2 M r}\right)^{-\epsilon}
\\
&\times &
\left\{1+qF_1(r)
P_2(\cos\theta)-(q+\epsilon q_*)\left[
2\ln\left(1+\frac{M^2\sin^2\theta}{r^2-2 M r}\right)
+
3F_2(r)\sin^2\theta\right]\right\}
+
{\cal O}(q^2,q_*^2)\ ,
\\
g_{\phi\phi} &=& r^2\sin^2\theta
\Big[1-qF_1(r)P_2(\cos\theta)\Big]
+
{\cal O}(q^2) \ ,
\end{eqnarray}
\end{subequations}
\end{widetext}
which is a generalized form
of the Erez-Rosen metric with external parameter
$q_*$ produced by the gravitational scalar field
where $P_2(\cos\theta) = (3\cos^2\theta-1)/2$
and the functions $F_1(r)$ and $F_2(r)$ are defined as

\begin{eqnarray}
F_1(r)&=&
3\left(\frac{r}{M}-1\right)
+
\left(\frac{3r^2}{2M^2}-\frac{3r}{M}
+1\right)
\ln\left(1-\frac{2M}{r}\right)\ ,
\\
F_2(r)&=&2+\left(\frac{r}{M}-1\right)
\ln\left(1-\frac{2M}{r}\right)\ .
\end{eqnarray}
%
%-------------------------------------------------------------------------
%-------------------------------------------------------------------------
%

\subsection{Test particle motion}
\begin{figure*}
	\includegraphics[width=\hsize]{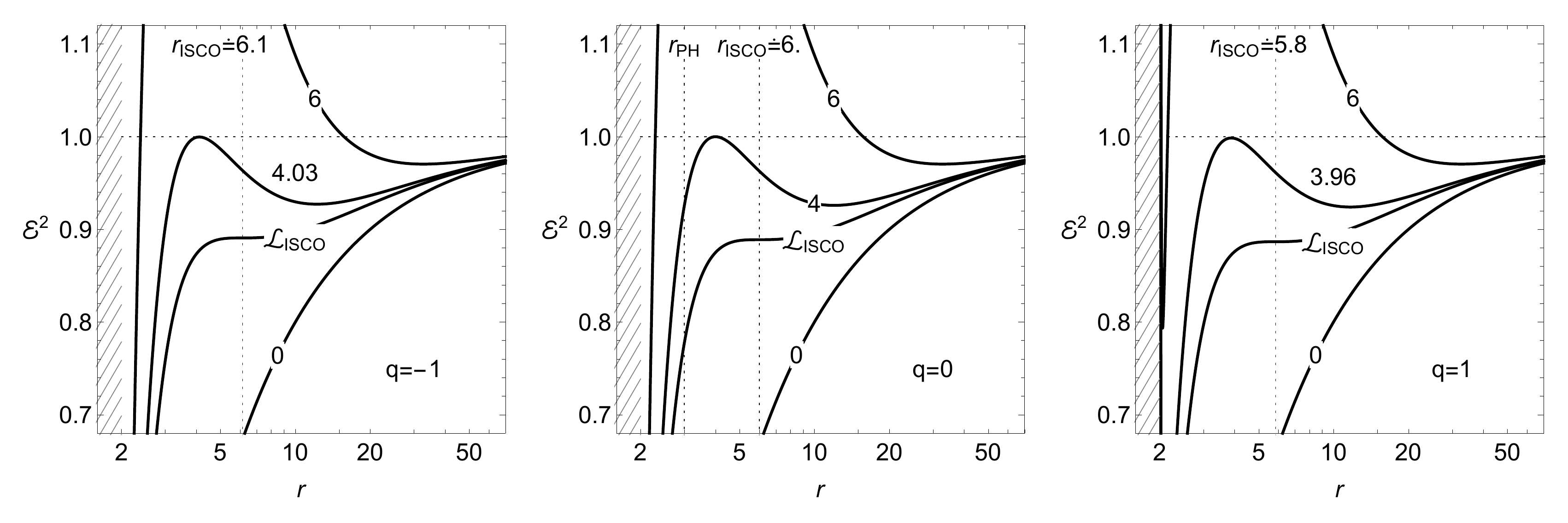}
	\caption{Radial profiles of effective potential
		in equatorial plane $V_{\rm eff}(r,\pi/2)$ for
		the various values of angular momentum $L$.
		In the plots the different values for $q$ parameter are used.\label{figVeffq}}
\end{figure*}

Consider the particle motion in the spacetime
metric (\ref{metric2}) with the linear term of
quadrupole momenta $q$ and $q_*$, Using the
equation of motion for the test particle
we can obtain the following effective potential

\begin{equation}
V_{\rm eff}(r)
=
\left(1-\frac{2M}{r}\right)
\left[1+\frac{{\cal L}^2}{r^2}
-
\frac{qF_1(r)}{2}\left(1
+
\frac{2{\cal L}^2}{r^2}\right)
\right]\ .
\label{Veff2}
\end{equation}
{Figure~\ref{figVeffq} draws radial dependence of the 
	effective potential for motion in the equatorial plane ($\theta=\pi/2$) 
	for the different values of the angular momentum, and 
	for three different values of quadruple moment $q$.}

In order to find the critical values of the energy and
the angular momentum one have to use the following conditions
\begin{equation}
{\cal E}^2 = V_{\rm eff}(r)\ ,\quad V_{\rm eff}'(r)=0\ ,
\end{equation}
and the solution of these equations for the energy
and the angular momentum can be found as
\begin{widetext}
\begin{eqnarray}
\label{Eq}
\nonumber
{\cal E}_{\rm ext}^2 &=& \frac{(r-2 M)^2}{r(r-3M)} -
q\Bigg[\frac{(r-M)(r-2M)
\left(6r^2-21Mr+19M^2\right)}{2Mr(r-3M)^2}
\\
&&+\frac{(r-2M)^2\left(6r^3-21Mr^2+23M^2r-6M^3\right)
}{4M^2r(r-3M)^2}\ln\left(1-\frac{2 M}{r}\right)\Bigg]
+ {\cal O}\left(q^2\right)\ ,
\\
\label{Lq}
\nonumber
{\cal L}_{\rm ext}^2 &=& \frac{M r^2}{r-3M} -
q\Bigg[\frac{r^2(r-M)\left(3r^2-9Mr+10M^2\right)}
{2M(r-3M)^2}
\\
&&
+\frac{r^2\left(3r^4-15Mr^3+30M^2r^2-26M^3r+6M^4
\right)}{4 M^2(r-3M)^2}\ln\left(1-\frac{2 M}{r}\right)\Bigg]
+ {\cal O}\left(q^2\right)\ .
\end{eqnarray}

The radius of ISCO can be found from the condition
$V_{\rm eff}''(r)\le 0$ in addition to expressions (\ref{Eq}) and (\ref{Lq})
which allow us to obtain the following nonlinear equation
\begin{eqnarray}
\label{ISCOq}
\nonumber
&&r-6M -
q\Bigg[\frac{12r^5-111Mr^4+364M^2r^3-501M^3r^2+214M^4r+54M^5
}{2M^2(r-2M)(r-3M)}
\\
&&+\frac{12r^5-99Mr^4+273M^2r^3-286M^3r^2+54M^4r+36M^5}
{4M^3(r-3M)}\ln\left(1-\frac{2 M}{r}\right)\Bigg]
+ {\cal O}\left(q^2\right)=0\ .
\end{eqnarray}
\end{widetext}
%

%-------------------------------------------------------------------------%
%-------------------------------------------------------------------------%

Obviously, it is difficult to get
analytical solution of the equation (\ref{ISCOq}).
Hereafter performing a careful numerical analysis
of expression (\ref{ISCOq}), one can find
that the radius of the ISCO  decreases with
increasing of the value of $q$ parameter.
In Table~\ref{Tab} a list
of numerical solutions for the radius of ISCO,
the energy and the angular momentum
of particles for different values of the
mass quadrupole moment is shown.
With the increase of the parameter $q$
radius of the ISCO to a gravitational object,
the values of the energy and the angular
momentum of particles decrease.
It means that the mass quadrupole moment $q$
sustain stability of particles circularly orbiting
around the black hole. One can conclude that
due to the mass quadrupole moment of the
black hole particles motion is more stable
than that in the Schwarzschild spacetime.
\begin{table}
\caption{\label{Tab} Dependence of the mass quadrupole
moment $q$ from the values of the radius of ISCO ($r_{\rm ISCO}$),
the critical energy (${\cal E}_{\rm ISCO}$) and angular momentum
(${\cal L}_{\rm ISCO}$) of the the test particles orbiting
around the black hole.}
\begin{ruledtabular}
\begin{tabular}{ccccc}
$q$ & ${r_{\rm ISCO}}/M$ & ${\cal L}_{\rm ISCO}/M$ & ${\cal E}_{\rm ISCO}$
\\
\hline
0   & 6.00000 & 3.46410 & 0.888889 \\
0.1 & 5.98552 & 3.46155 & 0.888684 \\
0.2 & 5.97090 & 3.45898 & 0.888478 \\
0.3 & 5.95616 & 3.45640 & 0.888269 \\
0.4 & 5.94127 & 3.45379 & 0.888057 \\
0.5 & 5.92624 & 3.45116 & 0.887844 \\
0.6 & 5.91107 & 3.44852 & 0.887627 \\
0.7 & 5.89575 & 3.44585 & 0.887408 \\
0.8 & 5.88027 & 3.44316 & 0.887187 \\
0.9 & 5.86464 & 3.44046 & 0.886963 \\
1.0 & 5.84884 & 3.43773 & 0.886736 \\
\end{tabular}
\end{ruledtabular}
\end{table}

\begin{figure*}
\includegraphics[width=\hsize]{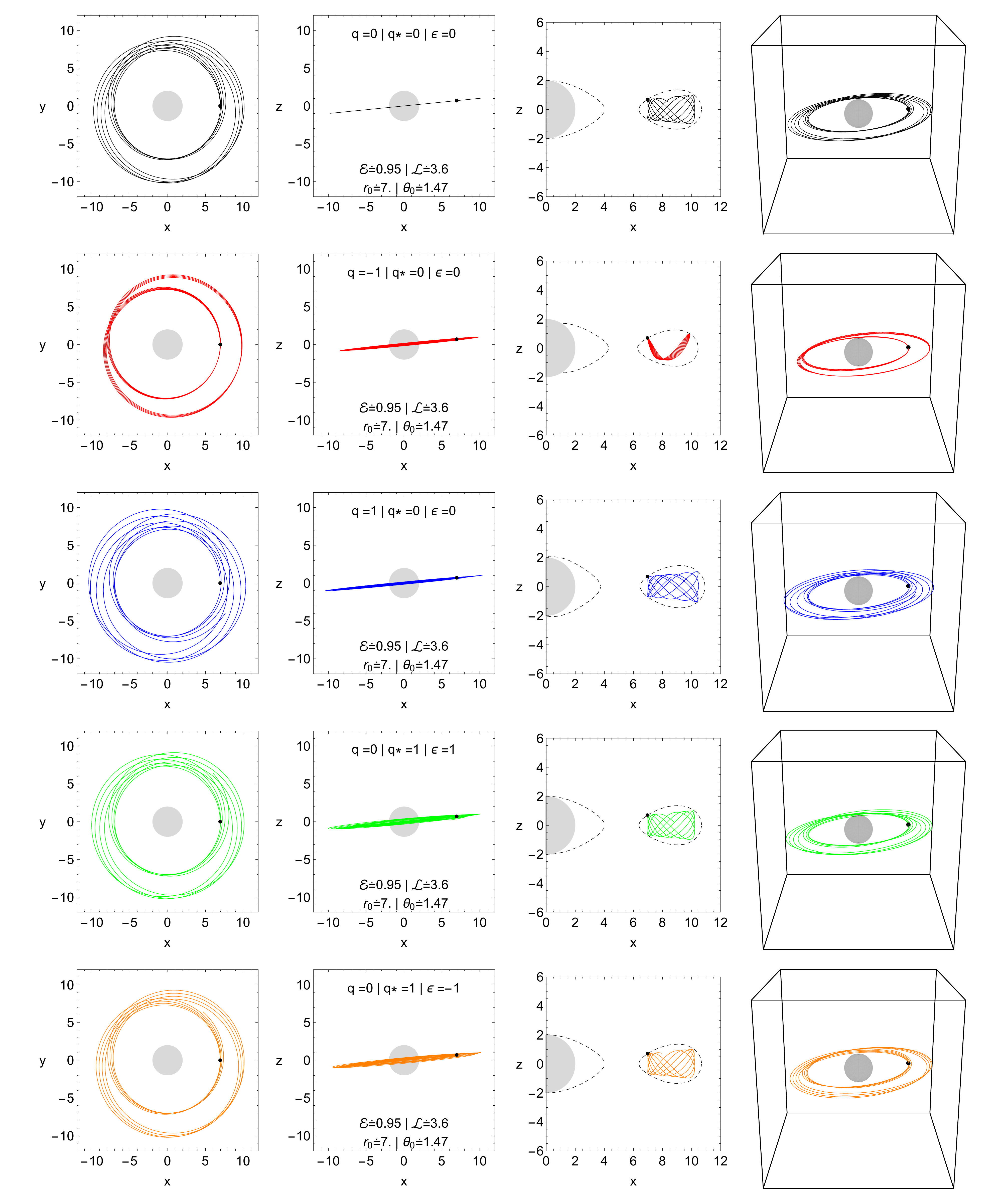}
\caption{{Test particle trajectories in the spacetime metric (\ref{metric2}) for the different values of 
	parameters $q$, $q_*$ and $\epsilon$. In first and second (including third) columns
	 particle trajectories in $x-y$ and $x-z$ planes are given, respectively,  
	while in the fourth column 3D $x-y-z$ pattern of 
	particle trajectory is shown.}\label{Trajectoryq}}
\end{figure*}

{The trajectories of the test particles 
in the spacetime of the generalized Erez-Rosen
metric (\ref{metric2}) at the several planes
for the different values of the parameters 
are shown in Fig.~\ref{Trajectoryq}.
The motion of the test particle becomes 
regular (not chaotic as in the Kerr 
spacetime) in the quadrupole moment metric.}

\begin{figure*}
\includegraphics[width=\hsize]{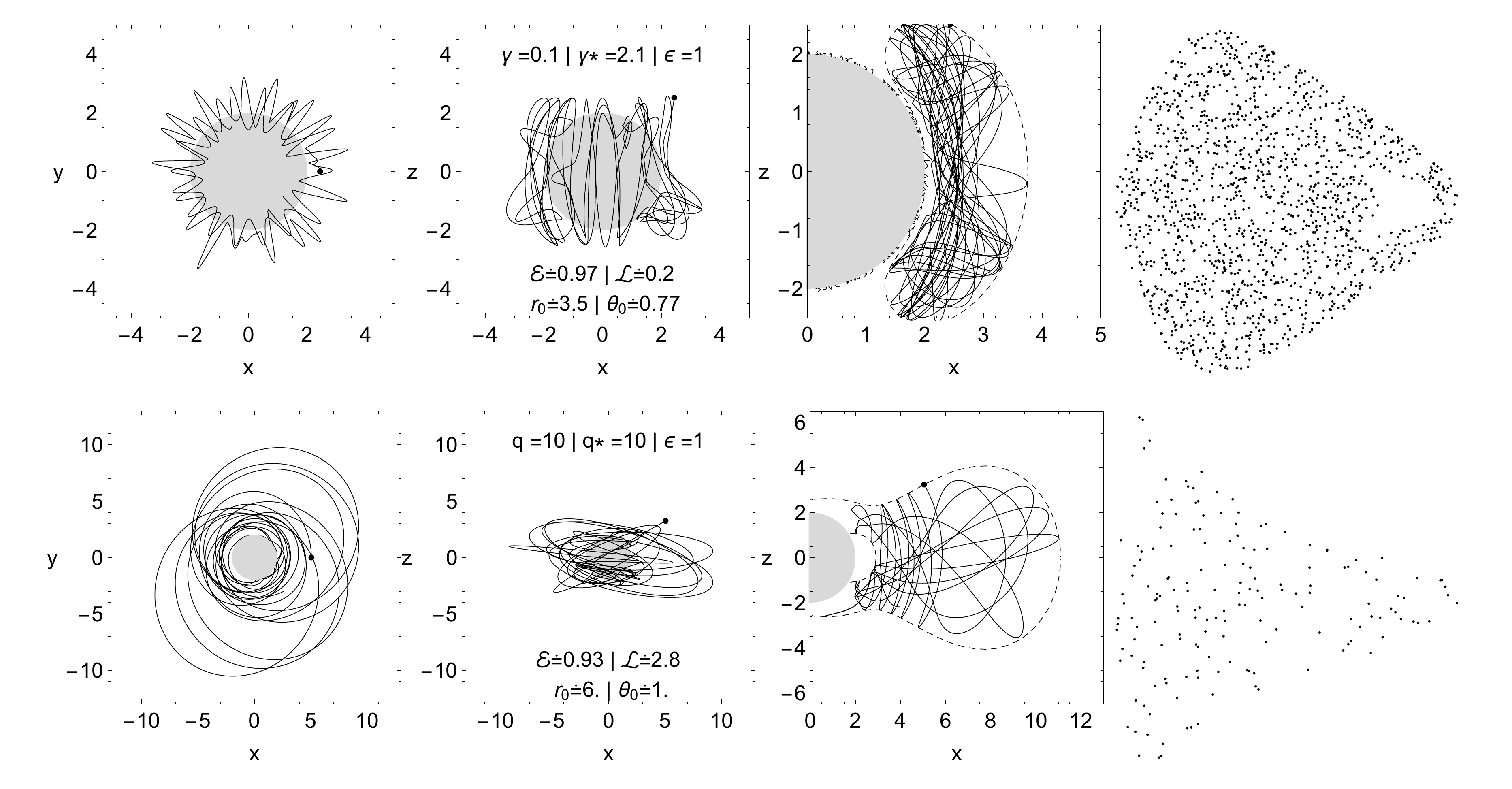}
\caption{{Chaotic trajectories of test particle 
		in several planes in background geometry decribed by (\ref{metric1}) and 
		(\ref{metric2}) when $\epsilon=1$.
		In first and second (including third) columns
		 particle trajectories $x-y$ and $x-z$ planes are given
		while in the fourth column phase-space diagram of 
		particle trajectory is shown.}\label{Trajectorygq}}
\end{figure*}

{It is interesting to study 
chaotic motion in the spacetime with deformation 
parameters $\gamma$, $\gamma_*$, $q$ and $q_*$. 
In order to check chaotic motion around the 
black hole we have used the general form of the 
spacetime metric which is given by the expressions (\ref{Phiq})-(\ref{Vq}).  
Numerical calculations show that the
trajectory of test particles become chaotic 
for large values of the $\gamma_*$, $q$, and $q_*$ 
parameters, as shown in Fig.~\ref{Trajectorygq}.}

%------------------------------------------------------------------------
%----------

%------------------------------------------------------------------------
%----------

\subsection{Energy conditions}

{Using the expression for the
energy-momentum (\ref{EMtensor}) one can
easily find the density and the components
of the pressure for scalar field defined in 
equation (\ref{Phisphere}) in the form}

\begin{widetext}
\begin{eqnarray}
\label{EMtensorq}
\nonumber
\rho &=&
P_\theta=P_\phi=-P_r
\\\nonumber
&=&
- \frac{\epsilon M^2}{2r^3(r-2M)}
\left(1+\frac{M^2\sin^2\theta}{r(r-2M)}\right)^{\epsilon}
\Bigg\{1+
q \left[2\ln\left(
1+\frac{M^2\sin^2\theta}{r(r-2M)}\right)
-F_1(r)P_2(\cos\theta)+
3F_2(r)\sin ^2\theta \right]
\\
&+&
q_*\left[2\epsilon\ln\left(1+\frac{M^2\sin^2\theta}{r(r-2M)}
\right)+3\epsilon F_2(r)\sin^2\theta
+1-\frac{6r}{M}+\frac{3r^2}{M^2}+\frac{3r}{2M}
\left(\frac{r}{M}-2\right)\left(\frac{r}{M}-1\right)
\ln\left(1-\frac{2M}{r}\right)\right]
\Bigg\}\ ,
\end{eqnarray}
\end{widetext}
{from equation (\ref{EMtensorq})
we can see that $\rho+P_r = 0$ which always
satisfies the NEC condition while the expressions 
for $\rho+P_{\theta,\phi}$
satisfy the NEC condition
only in the case when $\epsilon \le 0$
which corresponds to the phantom field.}

\section{Conclusion\label{Summary}}

In the present paper we have derived
axisymmetric and static solutions of the
Einstein field equations by taking into
account the effect of an additional self-gravitating scalar field. In particular we have presented an exact analytical solution of the combined Einstein equations for two different modified spacetime metrics which belong to the Weyl class of solutions as (i) the modified
$\gamma$-metric and (ii) the modified quadrupole moment metric. Obtained results can be summarized as follows:

\begin{itemize}
\item
We have studied the influence of the scalar field
in spacetime properties of axial-symmetric
and static vacuum solutions
of combined Einstein field equations which generalize the
Schwarzschild's spherically symmetric solution
to include $\gamma$, $\gamma_*$ and mass
quadrupole parameters $q$, $q_*$.
We have required that the scalar field is
axially symmetric, static and that
the solutions satisfy the
asymptotic flatness and curvature regularity.
We have obtained a generalized form of the 
$\gamma$ metric with additional the $\gamma_*$ and generalized form of the Eres-Rosen (quadrupole moment) metric which includes $q_*$ mass quadrupole produced by
 the self-gravitating scalar field.

\item
The analytical expressions for the
components of the energy-momentum tensor are
obtained for the self-gravitating scalar field.
Extensive analysis of the energy of scalar field
has shown that in the case of phantom field
($\epsilon=-1$) it satisfies the NEC while in the case of
gravitating scalar field ($\epsilon=1$) it does not
satisfy the NEC.

\item
We have studied the test particle motion in 
the spacetime of both generalized the 
$\gamma$-metric and the quadrupole moment
metric and  have probed $\gamma_*$ and $q_*$
parameters produced by the gravitating scalar field
into the test particle motion. The Hamilton-Jacobi
 equation of motion
for the test particle is chosen as in our preceding research
done in Ref.~\cite{Kolos15}.
Our analysis shows that $\gamma_*$ and $q_*$ parameters
do not contribute into the energy and angular momentum
of the test particle and consequently do not affect
particle trajectory at the equatorial plane.
{Consequently, the motion of the test
particle becomes regular (rather than chaotic)
in both generalized $\gamma$ and generalized
Erez-Rosen metrics.}

\item
We have presented the exact analytical expression
for the radius of the ISCO, the critical values of the
energy and the angular momentum of the test particles
in terms of the $\gamma$ parameter in the spacetime
of the $\gamma$-metric. It is shown that
for the range $\gamma\ge 1/\sqrt{5}$ of the 
values of the $\gamma$ parameter  the radius of
ISCO and the photon sphere increase.
For the range of the values of the $\gamma\ge 1$ 
we have found
that with increasing the $\gamma$ parameter the radius of
ISCO and photon sphere increase while for the range of the values
$1/\sqrt{5}\le\gamma\le 1$ they are small in comparison
with that in the Schwarzschild spacetime.
After performing numerical analysis of the 
equations of particle
motion in the spacetime of the generalized
quadrupole moment metric we have found that
the radius of ISCO  decreases with
an increase of the value of the $q$ parameter.
{It has been shown that the quadrupole
moment metric has circular orbits that are more 
strongly bounded when compared to that in 
the Schwarzschild metric.}

\end{itemize}

\section{acknowledgments}

The research of B.A. and B.T. is partially
supported by Grants No. VA- FA-F-2-008 and
No. YFA-Ftech-2018-8 of the Uzbekistan Ministry
for Innovation Development, and by the Abdus Salam
International Centre for Theoretical Physics
through Grant No. OEA-NT-01. B.A. would like
to acknowledge the support of the
German Academic Exchange Service DAAD for
supporting his stay at Frankfurt University.
M.K. acknowledges the Czech Science Foundation
under the Grant No. 16-03564Y. Z.S. acknowledges
the Albert Einstein Centre for Gravitational and
Astrophysics supported under the Czech Science
Foundation under the Grant No. 14-37086G.
The authors would like to thank Professor H.~Quevedo
for pointing out the methods of derivation of
new solutions of the Einstein equations
and Professor Y.~M.~Cho for suggesting we include
the scalar field in the solutions. B.A. acknowledges the support from Nazarbayev University Faculty Development Competitive Research Grants: ``Quantum gravity from outer space and the search for new extreme astrophysical phenomena'', Grant No. 090118FD5348.

\begin{appendix}
\section{The function $V(X,Y)$ \label{App1}}

The explicit form of the function $V(X,Y)$ in
the equation (\ref{SolV}) is given by~\cite{Quevedo89}
\begin{equation}
V(X,Y) = \sum_{l,n=0}^{\infty} (-1)^{l+n}
(q_lq_n+\epsilon p_lp_n)\Gamma^{\{ln\}}\ ,
\end{equation}
where
\begin{widetext}
\begin{eqnarray}
  \nonumber
  \Gamma^{\{ln\}} &=& \frac{1}{2}\ln\frac{X^2-1}{X^2-Y^2}
  +
  \left(\epsilon_n+\epsilon_l-2\epsilon_n\epsilon_l\right)
  \ln\frac{X+1}{X-1}
  \\\nonumber
  &+&
  (X^2-1)\Big[X\Big(A_{n,l} Q_n'(X)Q_l(X)+A_{l,n}Q_l'(X) Q_n(X)\Big)
  -C_{n,l}Q_n(X)Q_l(X)\Big]
  \\\nonumber
  &+&
  (X^2-1)\left[(1-\epsilon_n){\cal C}_l+\epsilon_l
  {\cal C}_{l+1}
  - \frac{\epsilon_n}{l+1}\Big(P_l(Y)-(-1)^l\Big)Q_l'(X)\right]
  \\\nonumber
  &+&
  (X^2-1)^2\Big[Q_l(X) {\cal B}_{l,n} -Q_l'(X) {\cal A}_{l,n}
  +
  \frac{1}{n+1} A_{l,n}Q_l'(X)   Q_n'(X)\Big]  \ ,  
  \end{eqnarray}
with
\begin{equation}
\nonumber
\epsilon_n =  
\left\{
\begin{array}{lcr}
1\ , \quad {\rm n= even\,\, integer}\ ,
\\
\\
0\ , \quad {\rm n= odd \,\, integer}\ ,
\end{array}
\right.
\end{equation}
and
\begin{eqnarray}
{\cal A}_{l,n} &=& \sum_{k=0}^{[n/2-1]}
\left(\frac{1}{n-2k+1}+\frac{1}{n-2k}\right)
A_{l,n-2k}Q_{n-2k}'(X)\ ,
\\
{\cal B}_{l,n} &=& \sum_{k=0}^{[n/2-1]}
\left(\frac{1}{n-2k-1}+\frac{1}{n-2k}\right)
B_{l,n-2k-1}Q_{n-2k-1}'(X)\ ,
\\
{\cal C}_{l,n} &=& \sum_{k=0}^{[n/2-1]}
\left(\frac{1}{n-2k-1}+\frac{1}{n-2k}\right)
\left[P_{n-2k-1}+(-1)^{n+1}\right]Q_{n-2k-1}'(X)\ ,
\\
A_{l,n} &=& \sum_{k=0}^{[n/2-1]} \sum_{s=0}^{\mu(n,l-2k-1)}
\frac{(2l-4k-1)K(l-2k-1,n,l)}{2(l+n)-4(k+s)-1}
\left(P_{l+n-2(k+s)}(Y)-P_{l+n-2(k+s+1)}(Y)\right)  \ ,
\\
B_{l,n} &=&  \sum_{j=0}^{[n/2-1]}\sum_{k=0}^{[n/2-1]}
\sum_{s=0}^{\mu(n-2k-1,l-2j-1)}
\frac{(2l-4j-1)(2n-4k-1)K(l-2j-1,n-2k-1,s)}{2(l+n)-4(j+k+s)-3}
\\\nonumber
&\times &
\left(P_{l+n-2(j+k+s)-1}(Y)-P_{l+n-2(k+l+s)-3}(Y)\right)\ ,
\\
C_{l,n} &=& B_{n+1,l} - (n+1)A_{n,l} \ ,
\end{eqnarray}
\end{widetext}
here a bracket $[Q]$ denotes an integer
part of quantity $Q$ and $\mu(a,b)={\rm min}(a,b)$,
the Clebsch-Gordon coefficients $K(l,n,k)$ are defined by
\begin{equation}
K(l,n,k) = \frac{2l+2n-4k+1}{2l+2n-2k+1}\frac{a_{l-k}
a_{n-k}}{a_{l+n-k}}
\ ,
\end{equation}
and
$$
a_k=\frac{(2k-1)!!}{k!} \ .
$$

\end{appendix}

\bibliographystyle{apsrev4-1}  %% BibTeX style
\bibliography{Ref}
\end{document}